# Giant magnetic anisotropy at nanoscale: overcoming the superparamagnetic limit

A. Hernando[1,*], P. Crespo[1], M. A. García[1], E. Fernández Pinel[1], A. Fernández[2] and S. Penadés[3]

[1] Instituto de Magnetismo Aplicado, RENFE-UCM-CSIC and Departamento de Física de Materiales UCM. P. O. Box 155, Las Rozas, Madrid 28230. Spain

[2] Instituto de Ciencia de Materiales de Sevilla CSIC-UNSE, Américo Vespucio s/n, 41092 Sevilla, Spain and Departamento de Quimica Inorgánica, Universidad de Sevilla, Spain

[3] Grupo Carbohidratos, Lab. of Glyconanotechnology IIQ-CSIC, Américo Vespucio s/n, 41092 Sevilla, Spain.

* Corresponding author: ahernando@renfe.es

**Abstract.**

It has been recently observed for palladium and gold nanoparticles, that the magnetic moment at constant applied field does not change with temperature over the range comprised between 5 and 300 K. These samples with size smaller than 2.5 nm exhibit remanence up to room temperature. The permanent magnetism for so small samples up to so high temperatures has been explained as due to blocking of local magnetic moment by giant magnetic anisotropies. In this report we show, by analysing the anisotropy of thiol capped gold films, that the orbital momentum induced at the surface conduction electrons is crucial to understand the observed giant anisotropy. The orbital motion is driven by localised charge and/or spin through spin orbit interaction, that reaches extremely high values at the surfaces. The induced orbital moment gives rise to an effective field of the order of $10^3$ T that is responsible of the giant anisotropy.

**Text**

Magnetism at nanoscale presents a set of surprising experimental results. Gambardella et *al.* found that single Co atoms deposited onto Pt surfaces show a remarkable magnetic anisotropy with an easy axis perpendicular to the surface [1]. The corresponding anisotropy constant was estimated to be close to 10 meV per Co atom, larger than that corresponding to the harder magnetic material (2 meV per Co atom for $SmCo_5$). For 1.4 nm thiol capped gold nanoparticles (NPs) a permanent magnetism was detected up to room temperature [2,3] and a similar behaviour had been previously reported for 2.4 nm palladium NPs [4,5]. For the case of gold the appearance of magnetism was completely amazing provided the diamagnetic character of bulk samples and the low value of its density of states at the Fermi level. However, what was also unexpected for Pd and Au is that NPs with size smaller than 2.4 nm could exhibit blocked magnetism at 300 K [2-5]. Moreover the thermal dependence of magnetization for Pd and Au NPs is very weak between 5 and 300 K. By assuming a first order kinetics for the relaxation of the magnetic moments and an attempt frequency factor equal to approximately $10^{10}$ s$^{-1}$ the anisotropy constant for a particle 2nm size with blocking temperature above 300 K should be at least of $10^9$ Jm$^{-3}$ that corresponds to approximately 0.4 eV per atom. That is indeed an enormous value compared not only to the normal values for the harder magnetic materials but also to the value reported in ref [1].

Recently, Carmeli et *al.* [6,7] found that thiol capped gold surfaces exhibit a giant paramagnetism with 50$\mu_B$ (Bohr magnetons) per atom and an easy axis also perpendicular to the surface, i. e., along the z axis. In order to analyse the giant values of the anisotropy one can better use the results obtained for thiol capped gold films where the easy axis is uniform over the whole surface and the local anisotropy merges macroscopically.

We have prepared Au films capped with Lewis conjugate onto glass substrates by template-stripped gold method with atomically flat surfaces [8]. Lewis conjugate, that links via a covalent bond to Au surface through sulphur similarly to thiols, is known to enhance self assembly effects [9].

Initially, the magnetic properties of the bare substrate were measured. As figure 1a shows, the substrates exhibit a paramagnetic behaviour at low temperature that turns to diamagnetic over 150 K. The magnetization curves resulted identical measuring with the magnetic field in plane and perpendicular to the surface. This is the typical behaviour from magnetic impurities that decrease with temperature. Figure 1b shows the magnetization curves measured at 5 K before and after chemisorption of the Lewis conjugate. When the field is applied in plane, the magnetization is the same that before chemisorption, however applying the magnetic field perpendicular to the surface, the magnetization curve resulted clearly different. Chemisorption induces a new contribution to the magnetization which can be obtained by subtracting the curves before and after chemisorption process. Repeating the procedure at different temperatures we found that the new contribution to the magnetization perpendicular to the surface induced by the chemisorption results independent of temperature (figure 1c) in the range 5-300 K. For in-plane field there is no difference between the substrate and the sample with the Lewis conjugate chemisorbed in this range of temperature (figure 1d).

From this curves with the magnetic field perpendicular to the surface and considering that for Au [111] surface there is $2 \cdot 10^{14}$ atoms /cm$^2$ [6], the magnetic moment per surface atom is about 100 $\mu_B$ per atom. Two questions arise from these results a) the origin of this anisotropy and b) the magnetization process provided that the coercivity of the magnetization curve along the easy axis is negligible in compared to the expected values of the anisotropy field.

Let us first show that the origin of the anisotropy is the same as the origin of the giant moments. In a recent paper [10] the authors have explained this enormous magnetic moment as due to orbital motion of the surface quasi-free electrons around the island of radius $\xi$ formed by self-assembled thiols bonding. The presence of localised charge or localised spin drives the induction of orbital momentum at the conduction electrons to minimize the spin-orbit interaction, $\alpha_r \hbar^2$. The quantum number $l_z$ that minimizes the kinetics and spin-orbit energy terms is easily obtained to be $l^*_z = m\xi^2\alpha_r s_z$ (m and $s_z$ hold for the mass and the third component of the spin of the free electron, respectively). The threshold strength of $\xi^2\alpha_r$, in order to trap an electron in the orbit, is that corresponding to $l^*_z=1$, i. e. the capture condition becomes $\xi^2\alpha_r > 2\cdot 10^{30}$ kg$^{-1}$ [11]. The strong spin-orbit interaction ($\alpha_r \hbar^2 = 0.4$ eV) at the gold surface [12,13] together with the large radius of the ordered islands (more that 50 nm) [14] drive the induction of giant orbital momenta. Furthermore, if instead of local electric charge the surface holds a localised spin, with z component $S_z$, the conduction electrons will also rotate around it by effect of the spin-orbit coupling. Thus, if the island consists of a localised ordered arrangement of both spin and charge, the orbital momentum that minimizes the energy should be given by $l^*_z = m\xi^2\alpha_r(S_z + s_z)$, where we assume $\alpha_r$ to be the experimental spin-orbit coupling strength measured for the particular surface. The spin-orbit interaction per atom that couples the localised charge and spin to the spin and orbital momenta of the conduction electron can be then written as:

$$H_{s\text{-}o}= \alpha_r \hbar^2 l_z(s_z+S_z) \qquad (1)$$

Where $l_z$ and $s_z$ are respectively the angular and spin momenta of the conduction electrons per atom comprised within the orbit [10]. The contribution to the surface magnetic moments comes from three sources (see Figure 2), a) the localised magnetic moment $g\mu_B J=g\mu_B(L+S)$, where g is the Landé factor and J the total angular momentum of a localised electron, hereinafter for the sake of clarity we will consider L=0, b) $\mu_B l_z$ is the orbital magnetic moment per atom induced on conduction electrons and c) $2\mu_B s_z$ that is the magnetic moment per atom associated with the spin of the conduction electrons with orbital momentum $l_z$. The total momentum depends on the sign of $\alpha_r$, but for the case of giant orbital moments the spin contribution is negligible. The rest of conduction electrons and those localised around the atomic cores gives rise to the well determined diamagnetic susceptibility of gold. Note that relation (1) points out the existence of an effective exchange coupling between the orbital moment and the spins of both orbital trapped and localised electrons.

A first remark is related to the constriction for l orientation. The conduction electrons rotate around localised charges and/or localised spins along orbits contained on the surface and consequently l can only have z component. Therefore, rotation of orbital magnetic moment is meaningless since it would mean that the electrons leave the metallic surface. However, according to (1), for any given $l_z$ its reversal does not change the energy provided a simultaneous reversal of $(s_z+S_z)$.

Since for thiol capped gold films the orbital moment is giant, its contribution is the only one that can be observed in the macroscopic magnetization. Therefore, independently of temperature, the measured anisotropy seems to be infinity given that the orbits are not allowed to rotate, as depicted in Figure 1-b. When the induced orbital momentum is of a few $\hbar$ units, its associated magnetic moment is of the order of those associated with the spins and both contribute to the measured macroscopic magnetization as is the case for NPs. The localised spins even though subjected to possible anisotropies independent of the orbital motion, are mainly blocked by the effective spin-orbit field, H*, given by (1).

$$\mu_0 H^* = \alpha_r \hbar^2 l_z / \mu_B \qquad (2)$$

This field strength $\mu_0 H^*$ is close to 1000 T for $\alpha_r \hbar^2 = 0.4$ eV, that is the case of spins localised on gold surfaces with $l_z=1$. Furthermore, for Au films the localised spins are blocked through an effective field $10^2$ times higher due to the giant value of $l_z$ that can approximate to $10^2$. Thus, for the case of gold in both forms films and NPs the enormous strength of H* ($10^5$ and $10^3$ T, respectively) indicates that $S_z$, $s_z$ and $l_z$ remain stiffly coupled under any applied field. This giant effective field can not induce any reversal of the magnetic moments since its direction reverses with them, but accounts, however, for the blocking of moments observed in Au NPs up to room temperature [15].

The second point deals with the magnetization process. For capped Au films subjected to fields applied along the z axis no hysteresis is observed, as depicted in Figure 1-a. The perpendicular magnetization tends to saturate at relatively low fields compared to the infinity value of the anisotropy constant inferred from the in-plane magnetization curve. This curve shows the reversible character of the magnetization process. The slope of the magnetization curve is a consequence of the demagnetizing field. The lack of hysteresis indicates that at zero applied field, the number of orbits with quantum number $l_z$ is equal to that with $-l_z$ and that transitions between them do not need to overcome any barrier. Consequently, a very weak magnetic field enables $l_z$ reversal.

The hysteresis observed for Au and Pd NPs can be explained as follows. The localised spins are also subjected to the local structural anisotropy with constant k per atom and with an easy axis that we assume perpendicular to the surface. Therefore, if the localised spin were isolated, they will reverse for an opposite field of strength $H = 2k/\mu_0\mu_B$. However, the applied field acting on S and trying to reverse it also acts on the orbital moment that is rigidly linked to S, through (1), as illustrated by Figure 2. Consequently the reversal will take place at a field

$$H_k = 2k/(\mu_0\mu_B(2s_z+2S_z+l_z)) \quad (3)$$

When $l_z$ is of the order of 100, as is the case of gold films, the reversal field is negligible, the curve does not present hysteresis, and only demagnetizing effects are observed. For NPs, as $l_z$ is of the order of unity [10], the reversal field is reduced but some hysteresis may be still observed. In the case of NPs only those magnetic moment comprised within an $\varphi$ angle around the field axis reverse. (For an applied reversal field $H_a$, it is verified $H_a\cos\varphi = H_k$)

It is the more important inference according to equations (1) and (2) that even though the structural anisotropy constant k could be of the order of the usual values for bulk materials, 0.1 meV per atom, the hysteresis does not disappears with temperature in NPs. The strong effective field, H*, eliminates the thermal fluctuations effect and the spin proceeds blocked up to temperatures of the order $k_BT=0.4$ eV or T=5000 K, independently of the NPs size, provided that $l_z$ is different to zero. The magnetic moments are blocked by H*, rather than by k, and can remain blocked up to above 300 K when the NPs size is even smaller than 1 nm. Consequently, the magnetization under constant applied field does not depend on temperature over a broad interval between 5 and 300 K.

In summary, at conducting surfaces, with strong $\xi^2\alpha_r$ product, localised electric charge or localised spins can trap electrons in orbits with angular momentum $l_z$. This orbital moment can not rotate but only reverse its value under the action of external magnetic fields. The localised spins are subjected to a strong effective magnetic field due to its coupling to $l_z$ via spin-orbit interaction. The reversal of the magnetic moment preserves the rigid coupling between spins and orbital moments. The structural local anisotropy acting on the localised spins gives rise to hysteresis that becomes negligible for giant $l_z$ values. Thermal motion does not affect the probability jumps over the anisotropy barrier as a consequence of the giant field H* acting on the spin. It remains to be elucidated the intrinsic mechanism underlying the reversal of $l_z$. These conclusions allow us to tailor samples at nanoscale with permanent magnetism up to temperatures above room temperature. This new type of magnetism is associated with surfaces where adequate capping and restrictions for orbital momentum rotation enable the induction of giant H*. Thus, high spin-orbit surfaces, with localised spins or electric charge distributions and well defined orbits, as has been shown for a few cases, are expected to exhibit amazing magnetic properties completely different to those well known for 3d and 4f elements.

## Acknowledgements

This work has been partially supported by the projects NAN2004-09125-C07-05 (Spanish Ministry of Education and Science) and 200560F0174 (Spanish Council for Scientific Research).

# Figure caption

**Figure 1**. Magnetization curves for (a) the bare substrate with H perpendicular to the surface; the same result was found for H parallel, (b) the bare substrate and the sample with H parallel and perpendicular at 5 K. (c) Difference between the magnetization curves for the substrate and the sample with H perpendicular to the surface and (d) H parallel to the surface.

**Figure 2**: Scheme of the different magnetic moments and the interaction controlling their orientation. The relative orientation of $S_Z$, $s_Z$ and $l_Z$ is fixed by $H^*$. The structural anisotropy acts only on $S_Z$ while the reversal magnetic applied field acts on all of them.

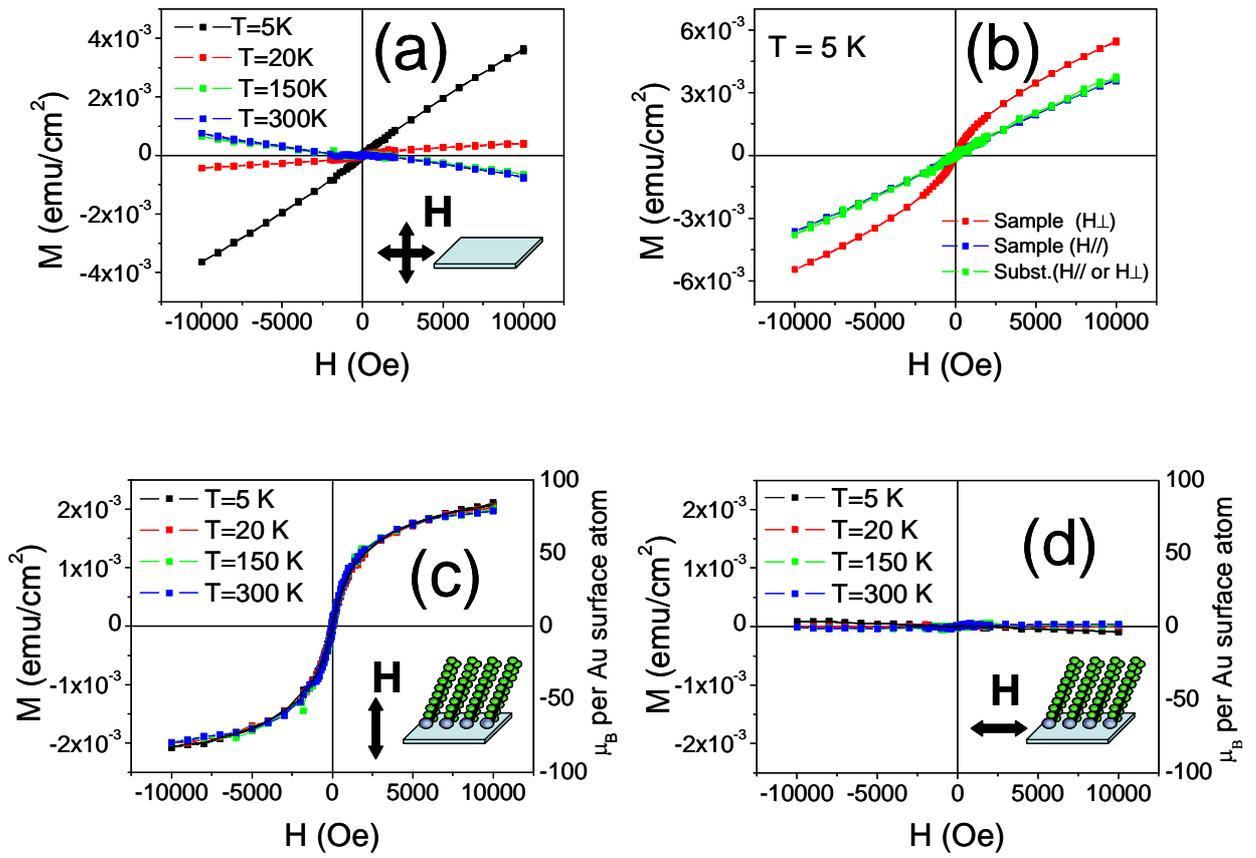

Figure 1 Hernando et al

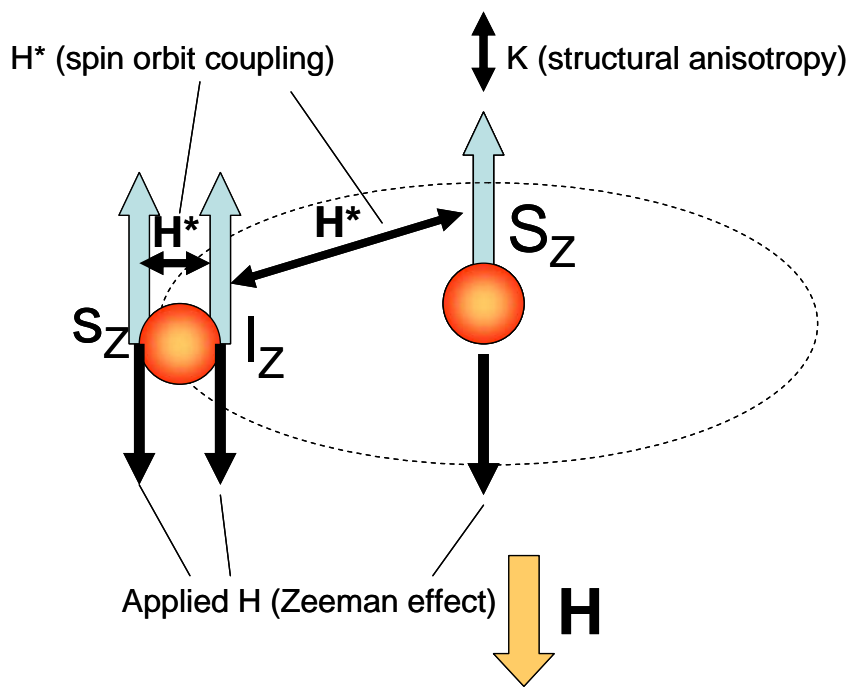

Figure 2 Hernando et al